\documentclass[preprint,12pt,authoryear]{elsarticle}
\usepackage{amsmath}
\usepackage{graphicx}
\usepackage{booktabs}
\usepackage{array}
\usepackage[T1]{fontenc}
\usepackage{lmodern}
\usepackage{url}
\usepackage[hidelinks]{hyperref}
\journal{Chemical Engineering Science}
\begin{document}
\begin{frontmatter}
\title{Flow-Field Design for Iron--Chromium Redox Flow Batteries Using the FlowBD-A1 Agent}
\author[a,b]{Zekun Jiang\fnref{eq}}
\author[a,b]{Shengwei Yuan\fnref{eq}}
\author[a]{Junliang Xiao}
\author[a]{Peihua Wu}
\author[a]{Siao Zhang}
\author[a]{Chunming Xu}
\author[a]{Yingchun Niu}
\author[a]{Tianhang Zhou\corref{cor}}
\ead{zhouth@cup.edu.cn}
\cortext[cor]{Corresponding author.}
\fntext[eq]{These authors contributed equally to this work.}
\affiliation[a]{organization={State Key Laboratory of Heavy Oil Processing, China University of Petroleum},
  city={Beijing}, postcode={102249}, country={China}}
\affiliation[b]{organization={College of New Energy Science and Engineering, China University of Petroleum},
  city={Beijing}, postcode={102249}, country={China}}

\begin{abstract}
In an iron--chromium redox flow battery, a better flow field must gain more in electrochemical performance than it costs in pumping power. FlowBD-A1 is an artificial intelligence (AI) agent for interdigitated flow-field design: it fits a Gaussian process (GP) to completed simulations and chooses the next designs by active learning or Bayesian optimization. We tested it by replaying 400 archived simulations, each described by channel depth and carbon-cloth utilization. Grouped five-fold validation gave a GP root-mean-square error of 0.737 percentage points and an $R^2$ of 0.891. Expected improvement and upper confidence bound recovered the highest archived efficiency (78.7748\%) in all 40 runs within 120 evaluations, compared with 12 of 40 for random search. Most cases above 78\% fell in a narrow utilization band (0.617--0.619), but the highest value lay at the sparsely sampled boundary (1.00). Uncertainty sampling found it within 25 evaluations in all 40 runs. Simulated fields show that this design's deep, wide channels and narrow rib drive the same flow through the electrode as neighboring geometries at the lowest inlet pressure. When simulations are expensive, these results favor exploring sparsely sampled regions early.
\end{abstract}

\end{frontmatter}

\section{Introduction}

In iron--chromium redox flow batteries, cell performance depends on electrochemical reaction, electrolyte transport and hydraulic losses~\citep{Sun2022,Zeng2016,Esan2020,Tang2014}. Electrode and channel configurations affect both reactant access and the pressure required to circulate electrolyte~\citep{Darling2014}. In an interdigitated flow field, the local pressure difference between adjacent inlet and outlet channels drives electrolyte through the porous carbon cloth. At a given circulation rate, the overall pressure drop determines the hydraulic power needed to sustain the flow. Channel width, spacing and depth change how flow is distributed through the electrode and how much pumping it requires. Better reactant delivery can reduce concentration losses, but any extra pumping power reduces the net gain. We therefore rank flow fields by voltage efficiency after pumping losses.

Autonomous experimentation combines evaluation tools with an algorithm that learns from completed measurements and chooses what to evaluate next~\citep{Hase2019}. Examples include a mobile robotic chemist, the Ada thin-film platform and the A-Lab for inorganic synthesis~\citep{Burger2020,MacLeod2020,Szymanski2023}. In simulation-based design, a numerical model supplies the feedback for the next decision. An AI agent uses its observations to choose actions for a given objective~\citep{Wooldridge1995}. Coscientist and ChemCrow use language models to plan tasks and select scientific tools~\citep{Boiko2023,Bran2024}. Other planners choose actions directly from a probabilistic surrogate; Atlas, for example, exposes Bayesian optimization through an ask--tell interface~\citep{Hickman2025}. FlowBD-A1 follows this route: after each batch of efficiencies arrives, its surrogate chooses the next batch of flow fields.

Active learning selects observations to improve knowledge of the response~\citep{Lookman2019,Kusne2020}, whereas Bayesian optimization uses an acquisition function to seek a stated objective~\citep{Jones1998,Snoek2012,Shahriari2016}. A GP supplies predictive means and uncertainty estimates that support both policies~\citep{Rasmussen2006}. Adaptive selection has been used to optimize battery charging protocols, electrolyte formulations and chemical reactions~\citep{Attia2020,Dave2022,Shields2021}. How quickly each policy reaches a good design depends on where the optimum lies relative to existing data, so we compare the two directly. In flow batteries, machine learning has been applied to materials and system operation, where data representation and validation remain open problems~\citep{Li2022}. Our previous work combined machine learning with a coupled flow and electrochemical model of interdigitated channels in iron--chromium flow batteries~\citep{Zhou2024}; the multiphysics model used here builds on that work. We represent each flow field by channel depth and carbon-cloth utilization. Depth measures the channel dimension normal to the electrode; utilization combines width and spacing, so distinct geometries can share a descriptor pair. Some archived cases with the same pair have different efficiencies. We keep all records and assign cases with the same pair to the same validation fold.

We test FlowBD-A1 on a fixed set of archived designs to see how the selection rule affects the number of evaluations needed to find the best flow field. We compare expected improvement (GP-EI), upper confidence bound (GP-UCB), uncertainty sampling (GP-AL), selection by the GP mean (GP-greedy) and random search. For a given run, each policy starts with the same cases and receives the efficiency of a new design only after selecting it. Predictive accuracy is assessed by grouped validation, and Shapley values describe the fitted response. Figure 1 shows the cell geometry and how the agent updates its choices. The released code also includes a COMSOL Multiphysics 6.3 adapter for new geometries.

\section{Methods}

\subsection{Simulation archive and physical descriptors}

The archive contains 400 records of channel depth $d$, carbon-cloth utilization $U_\mathrm{cc}$ and pump-loss-inclusive voltage efficiency. Depth spans 1.0--2.8 mm and dimensionless utilization spans 0.5--1.0. Recorded utilization values are used as supplied. Equation 1 defines the model input and objective.

\begin{equation}
x = (d, U_\mathrm{cc}), \qquad f(x) = \eta_\mathrm{sys}(x)
\label{eq1}
\end{equation}

The multiphysics model used to generate the archive couples open-channel Navier--Stokes flow, porous-electrode Brinkman flow, species transport and electrochemical reaction in a periodic flow-cell model~\citep{Zeng2016,Esan2020,Tang2014,Zhou2024}. Its archived geometry, mesh checks and physical-field panels are provided in Figures S1 and S2 of the Supporting Information. Replaying these responses lets us compare selection rules on the same set of simulations. Equation 2 accounts for pumping power at the stated operating point. $U_\mathrm{dis}$ and $U_\mathrm{ch}$ are discharging and charging voltages, $I_\mathrm{cell}$ is total cell current in amperes and $P_\mathrm{pump}$ is total pumping power. A current-density input requires conversion using the active area before electrical and hydraulic powers are combined. This metric is the operating-point voltage efficiency, and the archive stores its value for each record.

\begin{equation}
\eta_\mathrm{sys} = \frac{U_\mathrm{dis} I_\mathrm{cell} - P_\mathrm{pump}}{U_\mathrm{ch} I_\mathrm{cell} + P_\mathrm{pump}} \times 100\%
\label{eq2}
\end{equation}

The 400 rows occupy 329 exact descriptor coordinates. Twenty-five coordinates contain 96 records; the largest within-coordinate response range is 1.856 percentage points. All records are retained in the primary analysis. Exact coordinates stay together during predictive validation, and a fitted observation-noise term represents the within-coordinate response variation. A separate benchmark averages responses within each coordinate. Post-hoc grouping-precision and boundary-point diagnostics are also reported. Model checks retained with the archive include a polarization comparison with literature data at 80 mA\,cm$^{-2}$ and a specific flow rate of 5 mL\,min$^{-1}$\,cm$^{-2}$. In the mesh study, the inlet pressure stabilizes near 19,440 elements. The archived objective is evaluated at a state of charge (SOC) of 0.5.

\subsection{FlowBD-A1 architecture and evaluation}

FlowBD-A1 uses a GP model and a selection rule to decide which candidates to evaluate. PoolOptimizer stores observed responses, pending candidate identifiers, kernel parameters, feature scaling, acquisition settings and random-number state; run\_closed\_loop enforces the evaluation budget. Before each adaptive decision, the agent fits the GP to revealed observations and scores the remaining candidates. The user sets the objective, candidate set and budget; during the run, the agent selects candidates and updates its model without further input. Figure 1a links the interdigitated flow plates, porous carbon-cloth electrodes and membrane to the design descriptors. When configured for COMSOL (Figure 1b,c), the evaluator sets the geometry, builds the mesh, solves the coupled multiphysics model and extracts the electrical and hydraulic outputs needed for the objective. The agent calls ask to obtain candidate identifiers, invokes the evaluator, and passes completed observations to tell before the next GP fit (Figure 1d). The same ask and tell calls could be exposed through the Model Context Protocol (MCP)~\citep{Avila2025}.

We benchmark FlowBD-A1 using archived responses. The COMSOL adapter uses the same interface and requires a model file, parameter names, scalar output expressions and the full geometry for each candidate. The supplied examples invoke the agent and evaluator directly from Python. Responses remain hidden from the agent until their candidate is selected. The online adapter requests discharge and charge voltages in V, total cell current in A and total pumping power in W through explicit MPh evaluation units~\citep{MPh}. It requires exactly one finite scalar per output at the configured solution point; dataset and inner/outer solution selectors may be specified. A pumping-power expression must follow the model's total-flow and pump-efficiency convention. The declared objective uses the same pumping-power term for charge and discharge.

Before starting a solver, the example checks the model path, candidate descriptors, actual geometry columns, parameter mappings and output units. The evaluator logs solver errors and invalid responses and stops the run without substituting an objective value. Successful evaluations are cached with model/configuration and geometry checks; checkpoints preserve pending candidates so that interrupted batches can reuse completed evaluations.

\begin{figure}[htbp]
\centering
\includegraphics[width=\textwidth]{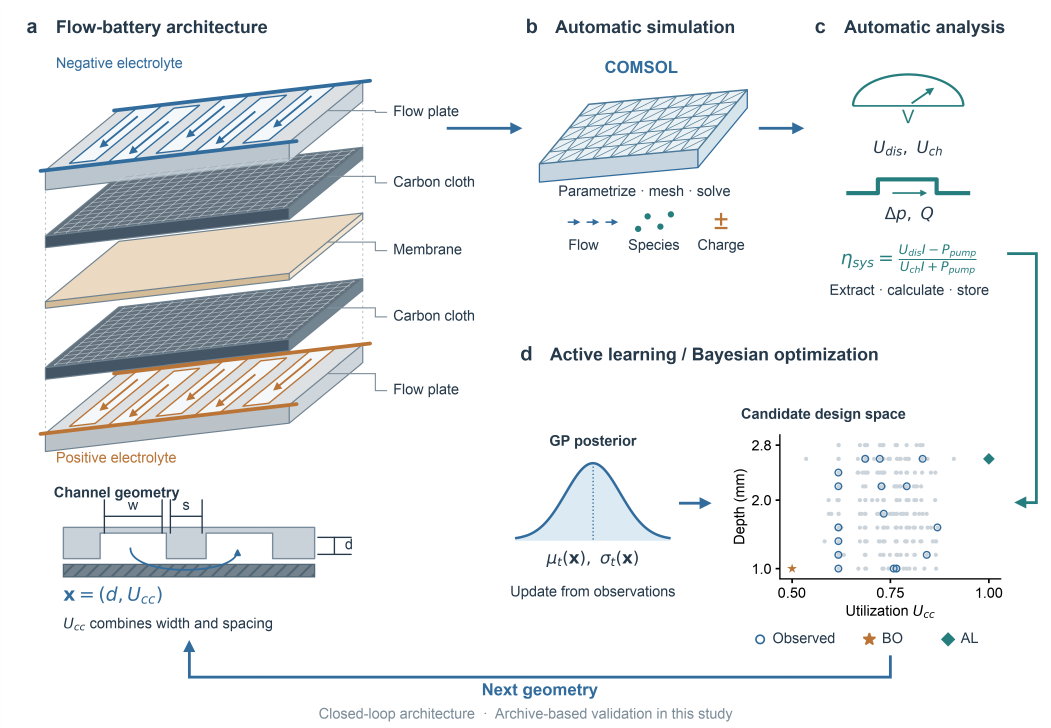}
\caption{FlowBD-A1 for interdigitated flow-field design. a, Exploded flow-cell assembly and a local channel section connect the interdigitated plates, carbon-cloth electrodes and membrane to the descriptors $d$ and $U_\mathrm{cc}$. b,c, For new geometries, the COMSOL evaluator constructs the geometry and mesh, solves coupled flow, transport and electrochemical equations, extracts electrical and hydraulic outputs, and calculates the pump-inclusive objective (shown as a fraction; multiply by 100 for percent). d, The agent updates the GP with the returned efficiencies. Active learning selects candidates by posterior uncertainty. Bayesian optimization uses EI or UCB. The map shows original candidate coordinates, the seed-0 initial observations, and the actual first single-candidate selections for GP-EI (BO) and GP-AL from the same 20 responses. Gray points denote the complete candidate pool. The battery, mesh, instruments and posterior-density icon are schematic; quantitative results use archived responses.}
\label{fig1}
\end{figure}

\subsection{Gaussian-process learning and acquisition}

The GP uses a Mat\'{e}rn 5/2 covariance with separate length scales for depth and utilization, an amplitude factor and additive white noise. Known candidate bounds scale each feature to the unit interval. Output centering and scaling use the currently observed responses only. Kernel hyperparameters are optimized by log marginal likelihood before adaptive batches 1, 5, 9 and so on, after the initial responses are available. A GP is fitted with the current observations at each intervening batch, retaining the previous kernel parameters. Fitting and acquisition use revealed responses only. For a candidate $x$, the GP supplies a posterior mean $\mu_t(x)$ and latent standard deviation $\sigma_t(x)$~\citep{Rasmussen2006}. Observation-noise variance is removed from the prediction covariance when constructing acquisition scores. GP-UCB maximizes equation 3 with $\kappa = 2.0$. GP-EI uses equation 4 with improvement offset $\xi = 0.01$ percentage points, $\Delta = \mu_t(x) - f_\mathrm{best} - \xi$ and $z = \Delta/\sigma_t(x)$. $\Phi$ and $\phi$ denote the standard-normal distribution and density functions~\citep{Jones1998,Snoek2012,Srinivas2010}.

\begin{equation}
a_\mathrm{UCB}(x) = \mu_t(x) + \kappa\, \sigma_t(x)
\label{eq3}
\end{equation}

\begin{equation}
a_\mathrm{EI}(x) = \Delta\, \Phi(z) + \sigma_t(x)\, \phi(z)
\label{eq4}
\end{equation}

Selecting a batch before responses arrive requires accounting for pending evaluations; GP bandit methods provide one formal treatment of this problem~\citep{Desautels2014}. FlowBD-A1 builds each batch one candidate at a time. After each choice, it updates the covariance for the remaining candidates using a temporary response equal to the posterior mean. The mean predictions are unchanged during batch construction. The next GP fit uses the actual responses. GP-EI applies expected improvement separately at each selection within the batch. The GP-AL control maximizes latent standard deviation, GP-greedy maximizes the posterior mean and random search samples the remaining candidates uniformly without replacement.

\subsection{Validation and paired optimization protocols}

Predictive validation uses five shuffled GroupKFold splits with seed 20261110. Exact duplicate descriptor pairs stay in one fold, so no shared coordinate appears in both training and test data; this avoids a form of leakage described by \citet{Kapoor2023}. Ten regressors (GP, gradient boosting, random forest, AdaBoost, decision tree, support vector regression, nearest neighbors, ridge, elastic net and lasso) use identical folds. The models are implemented in scikit-learn~\citep{Pedregosa2011}, and their fixed configurations are recorded in the code. Out-of-fold RMSE and $R^2$ are computed across the same 400 held-out predictions. GP interval coverage includes observation noise and is distinct from the latent uncertainty used by acquisition. Model configurations, acquisition parameters and benchmark budgets are fixed in advance and listed in the code and Supporting Information. Candidate selection in each replay depends only on the responses revealed in that replay.

The main optimization protocol uses 40 paired random initializations of 20 records, batches of 5 and a total budget of 120 revealed records. Each seed gives every method the same initial records. Candidate order is independently shuffled because the source table is sorted by efficiency, and record identifiers are excluded from the model features. Two additional protocols use 20 paired seeds each. One starts from 100 initial records with batch size 20 and budget 300; the other uses the 329-coordinate mean-response pool with the main protocol's initialization, batch size and budget.

Performance is assessed using the best revealed response, exact-maximum recovery, evaluation count to that maximum, and budget-averaged simple regret. Regret is the pool maximum minus the best revealed response. Because results arrive in batches, regret remains at its left endpoint value until each batch is completed. Budget-averaged regret is the area under this step curve divided by the post-initialization budget span. The pool maximum is used only for scoring. Runs that do not reach the target by the budget limit are right-censored. A 0.05-percentage-point tolerance provides a separate practical target.

\subsection{Exact two-feature interpretation}

A separate GP fitted to the complete archive supplies descriptive response maps and Shapley explanations; it is used only for interpretation. With two features, interventional Shapley contributions can be evaluated exactly by averaging over the two possible feature orders~\citep{Lundberg2017}. A fixed empirical background of 100 source rows defines the reference expectation. For depth, equation 5 uses $v_d = \mathrm{E}[f(d,U_b)]$, $v_u = \mathrm{E}[f(d_b,U)]$ and $v_0 = \mathrm{E}[f(d_b,U_b)]$; the utilization contribution follows by exchanging the two features. Additivity is checked numerically for all 400 records.

\begin{equation}
\phi_d = \frac{(v_d - v_0) + (f(x) - v_u)}{2}
\label{eq5}
\end{equation}

The main figure displays posterior predictions only at the 329 observed descriptor coordinates; continuous model grids are retained as diagnostic output. Shapley values are plotted at original feature coordinates without jitter. They describe the GP mean under interventional feature mixing and the stated background distribution. Feature dependence can change the interpretation of Shapley explanations~\citep{Aas2021}, so the values are interpreted as properties of the fitted model within the sampled domain.

\subsection{Data provenance and computational reproducibility}

The source CSV is sorted by response; source IDs are kept as provenance labels, and the candidate pool is independently shuffled with seed 20261110 before optimization. Initial subsets use seed 30000 plus the paired-run index; acquisition tie-breaking uses that index. A separately stored cumulative response table is used to cross-check the archived responses. The accompanying code provides a standalone Python example with FlowBD-A1, the archive evaluator, ten-model validation, all four benchmark protocols, an explicit-unit COMSOL adapter, reference results and automated checks. The quick example runs without COMSOL. Independent verification recomputes stored predictive scores, matches each queried response to its source pool, reconstructs completed-batch trajectories, checks censoring and regret integration, and tests Shapley additivity. Separate interface tests use mocked solver objects to check software behavior.

\section{Results and Discussion}

\subsection{Descriptor representation and predictive validation}

Archived efficiency ranges from 69.8111\% to 78.7748\%. Of the 38 records above 78\%, 31 have utilization between 0.617 and 0.619. Records 398--400 are isolated high-response cases at utilizations 0.6667, 0.50 and 1.00 (Figure 2a). The maximum is at depth 2.6 mm and utilization 1.00, 0.03970 percentage points above the second-highest response. The same three values appear in the separately stored cumulative response table, so they are retained as archive entries rather than treated as outliers.

The GP gives the lowest pooled out-of-fold RMSE among the ten specified regressors: 0.737 percentage points, with $R^2$ = 0.891 (Figure 2b,c; Table 1). Gradient boosting gives RMSE 0.781 and $R^2$ = 0.877; random forest gives RMSE 0.803 and $R^2$ = 0.870. Fold-specific scores vary across partitions (Figure 2c). The nominal 95\% observation-predictive intervals cover 379 of 400 held-out responses (94.75\%; Figure 2d), whereas nominal 50\% intervals cover 62.75\%. Coverage is close to nominal at 95\% and conservative at 50\%. Absolute held-out errors for records 398, 399 and 400 are 4.826, 4.402 and 4.279 percentage points; records 398 and 399 also lie outside their nominal 95\% intervals (Figure 2b). Record 400 remains inside its interval because its predictive standard deviation, 2.42 points, is nearly five times the median value. Records 7 and 8 are overpredicted by 4.110 and 4.126 points. Like records 399 and 400, they lie near the ends of the utilization range (0.50--0.54 and 0.91--1.00), where few training points constrain the GP.

\begin{table}[htbp]
\centering
\caption{Held-out predictive performance on identical grouped folds.}
\label{tab1}
\small
\begin{tabular}{lccc}
\toprule
Model & RMSE (pp) & MAE (pp) & R$^{2}$ \\
\midrule
GP & 0.7368 & 0.4196 & 0.8906 \\
GB & 0.7812 & 0.4089 & 0.8770 \\
RF & 0.8028 & 0.4016 & 0.8701 \\
SVR & 0.8143 & 0.5142 & 0.8663 \\
AdaBoost & 0.8380 & 0.4897 & 0.8584 \\
kNN & 0.8554 & 0.5102 & 0.8525 \\
DT & 0.8951 & 0.4602 & 0.8385 \\
Lasso & 1.0640 & 0.5718 & 0.7718 \\
Elastic net & 1.0645 & 0.5735 & 0.7716 \\
Ridge & 1.0649 & 0.5715 & 0.7714 \\
\bottomrule
\end{tabular}

\vspace{4pt}
\begin{minipage}{\linewidth}\footnotesize Table 1 pools all 400 held-out predictions for each model, so its RMSE values differ from an unweighted average of the five fold RMSEs. GB, gradient boosting; RF, random forest; SVR, support vector regression; kNN, k-nearest neighbors; DT, decision tree; pp, percentage points.\end{minipage}
\end{table}

\begin{figure}[htbp]
\centering
\includegraphics[width=\textwidth]{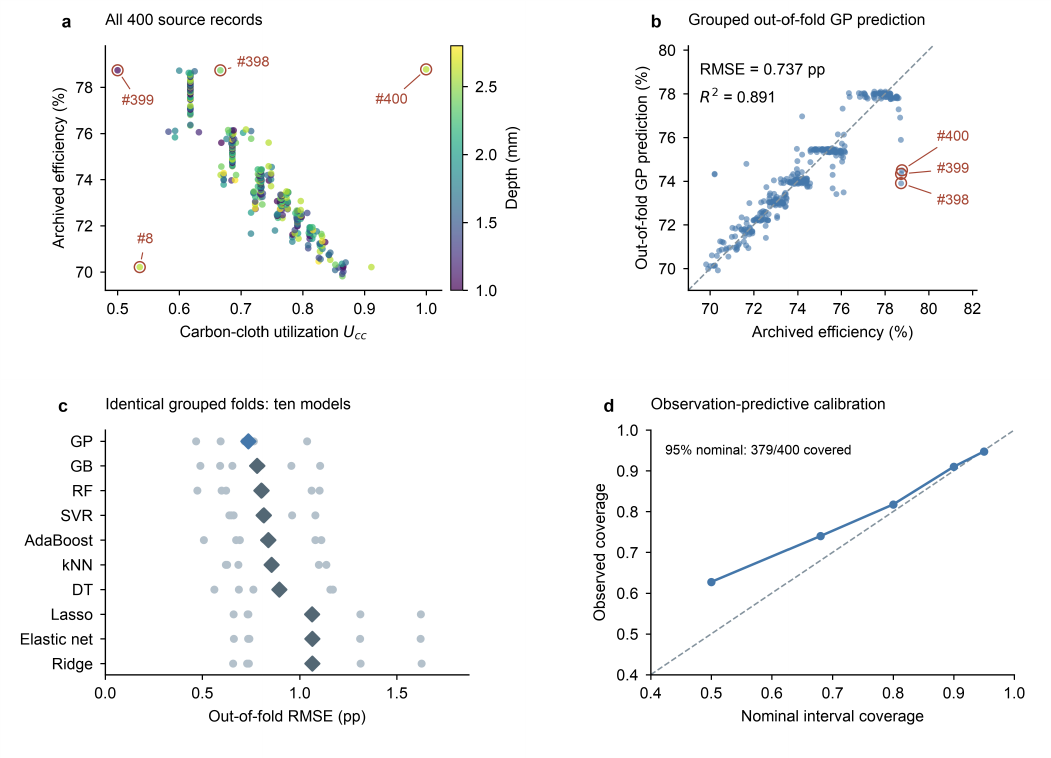}
\caption{Archive structure and grouped predictive validation. a, All 400 archived efficiencies at their original utilization coordinates, colored by channel depth. Labels identify source records 8 and 398--400. b, GP predictions from five-fold validation grouped by exact descriptor pairs; labels in the empty right margin identify the three highest-response records. Absolute prediction errors for records 398, 399 and 400 are 4.826, 4.402 and 4.279 percentage points (pp), respectively. The dashed line denotes identity. c, Fold-specific RMSEs (dots) and pooled out-of-fold RMSEs (diamonds) for ten regressors using the same folds. d, Descriptive coverage of GP observation-predictive intervals across 400 held-out responses. All source records are shown.}
\label{fig2}
\end{figure}

\subsection{Bayesian optimization and active learning performance}

GP-EI and GP-UCB recover the archived maximum in all 40 runs within the 120-evaluation budget, compared with 12 of 40 runs for random search (Figure 3; Table 2). GP-EI reaches 90\% recovery by evaluation 30 and 100\% by evaluation 35; GP-UCB reaches full recovery by evaluation 45. Median recovery counts are 30 for GP-EI and 32.5 for GP-UCB, including the 20 initial observations. Random search does not reach a median recovery count within the budget. GP-AL recovers the maximum in every run by evaluation 25, that is, within its first adaptive batch. This result follows from where the maximum lies. Record 400 sits at the upper utilization bound, where the GP is least constrained, and the first variance-maximizing batch selects it whenever the initial records lack it. GP-EI weighs that uncertainty against predicted efficiency. Its first batch more often reaches record 399 at the lower bound (29 of 40 runs), and in 25 runs record 400 follows one or two batches later. GP-greedy succeeds in 39 of 40 runs and has a median recovery count of 95; because it ignores uncertainty, it reaches the boundary maximum late.

Budget-averaged regret complements exact recovery by accounting for intermediate performance. Median values are 0.1388 percentage points for random search, 0.0582 for GP-greedy, 0.0243 for GP-AL, 0.0248 for GP-UCB and 0.0243 for GP-EI (Figure 3c). GP-AL and GP-EI share a median because both usually reach the maximum within one or two batches of the same initial records. Seed by seed, GP-EI never has lower regret than GP-AL (median paired difference 0.0020 points). The median paired reduction for GP-EI relative to random search is 0.1053 points, with a 95\% paired-seed bootstrap interval of 0.0554--0.2122 from 10,000 resamples. The paired effect is the median of seed-wise differences.

\begin{table}[htbp]
\centering
\caption{Main-protocol recovery and evaluation allocation.}
\label{tab2}
\small
\begin{tabular}{l*{4}{>{\centering\arraybackslash}p{0.17\linewidth}}}
\toprule
Policy & Exact best & Within 0.05 pp & Median count to exact best & Median regret (pp) \\
\midrule
Random & 12/40 & 26/40 & Not reached & 0.1388 \\
GP-greedy & 39/40 & 40/40 & 95 & 0.0582 \\
GP-AL & 40/40 & 40/40 & 25 & 0.0243 \\
GP-UCB & 40/40 & 40/40 & 32.5 & 0.0248 \\
GP-EI & 40/40 & 40/40 & 30 & 0.0243 \\
\bottomrule
\end{tabular}

\vspace{4pt}
\begin{minipage}{\linewidth}\footnotesize Median evaluation counts include the 20 initial records and are taken over all 40 runs, with unsuccessful runs right-censored at the budget. The random-search median is not reached because fewer than half of its runs recover the maximum. In one seed the initial records already contain the maximum, so every policy recovers it at evaluation 20. Budget-averaged regret covers only the post-initialization interval.\end{minipage}
\end{table}

\begin{figure}[htbp]
\centering
\includegraphics[width=\textwidth]{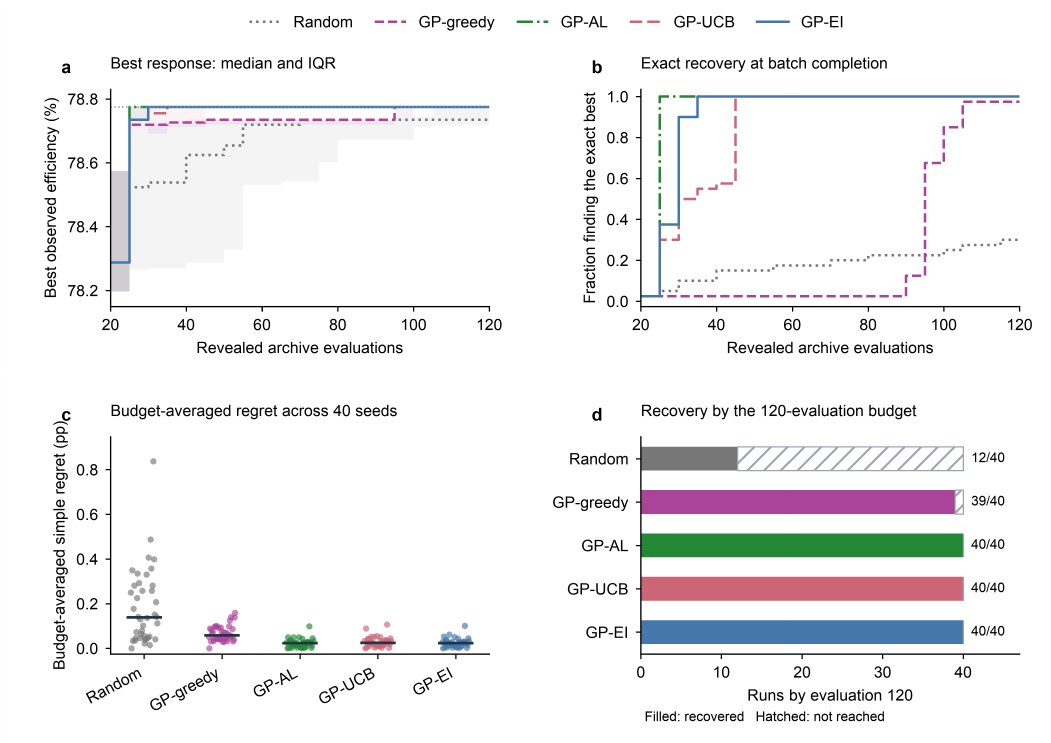}
\caption{Paired hidden-response replay across 40 shared random initializations, with 20 initial records, batches of five and a total budget of 120. a, Median best archived efficiency with the interquartile band across seeds. b, Fraction of runs recovering the exact pool maximum. Curves change only when a batch is completed. c, Budget-averaged simple regret for each seed; horizontal offsets separate seeds and horizontal marks denote medians. Regret uses the left endpoint value between completed batches and is divided by the post-initialization budget span. d, Numbers of runs recovering the maximum or remaining unsuccessful at the budget limit. GP-AL maximizes the latent posterior standard deviation (SD); GP-greedy maximizes the posterior mean.}
\label{fig3}
\end{figure}

\subsection{Learned response structure and attribution}

Across the observed coordinates, the full-data GP varies mainly with utilization (Figure 4a). Figure 4b shows its latent posterior uncertainty at the same points. The fitted depth length scale reaches its upper bound of 5 normalized units, whereas the utilization length scale is approximately 0.0226. Mean absolute Shapley contributions are 1.744 percentage points for utilization and 0.013 for depth (Figure 4c,d). Vertical columns in the depth plot arise from the discrete sampled depth levels. The small depth contribution means that, over 1.0--2.8 mm, the GP finds no depth trend that stands out from the scatter among records sharing a descriptor pair. Channel depth nonetheless affects hydraulic resistance and the pressure distribution that drives flow through the electrode. Utilization locates the main group of high-efficiency designs, but the best archived design, at utilization 1.00, falls outside it (Figure 2a).

The simulated fields in Figure S2 show how the best design improves flow relative to its neighbors. The archived model record gives record 400 a channel depth of 2.6 mm, a channel width of 2.6 mm and a channel spacing (rib width) of 2.0 mm. Figure S2 compares it with five geometries in which only the depth (1.6 or 1.0 mm) or only the width (2.0, 1.6 or 1.0 mm) is reduced. In all six, electrolyte passes from the inlet to the outlet channel through the electrode under the rib. There the velocity peaks at 0.016--0.017 m\,s$^{-1}$, and the local current density is an order of magnitude higher than under the channels (Figure S2h,i). The geometries differ in the pressure needed to drive this flow (Figure S2g). The 2.6-2.6-2.0 geometry needs the lowest inlet pressure, 1130 Pa, and nearly all of it is spent across the rib. With a 1.0-mm depth or a 1.0-mm width, the inlet pressure rises to 1240--1250 Pa, and 140 Pa is still needed to carry electrolyte along the outlet channel. Reducing the depth from 2.6 to 1.6 mm raises the inlet pressure by less than 40 Pa, consistent with the small depth contribution in Figure 4d. Deep, wide channels carry electrolyte along the flow field with little pressure loss, so the applied pressure goes into convection through the reacting electrode under the rib, and the 2.0-mm rib keeps that path short. The best design therefore supplies the electrode as well as its neighbors at a lower pumping cost, which raises its voltage efficiency after pumping losses.

\begin{figure}[htbp]
\centering
\includegraphics[width=\textwidth]{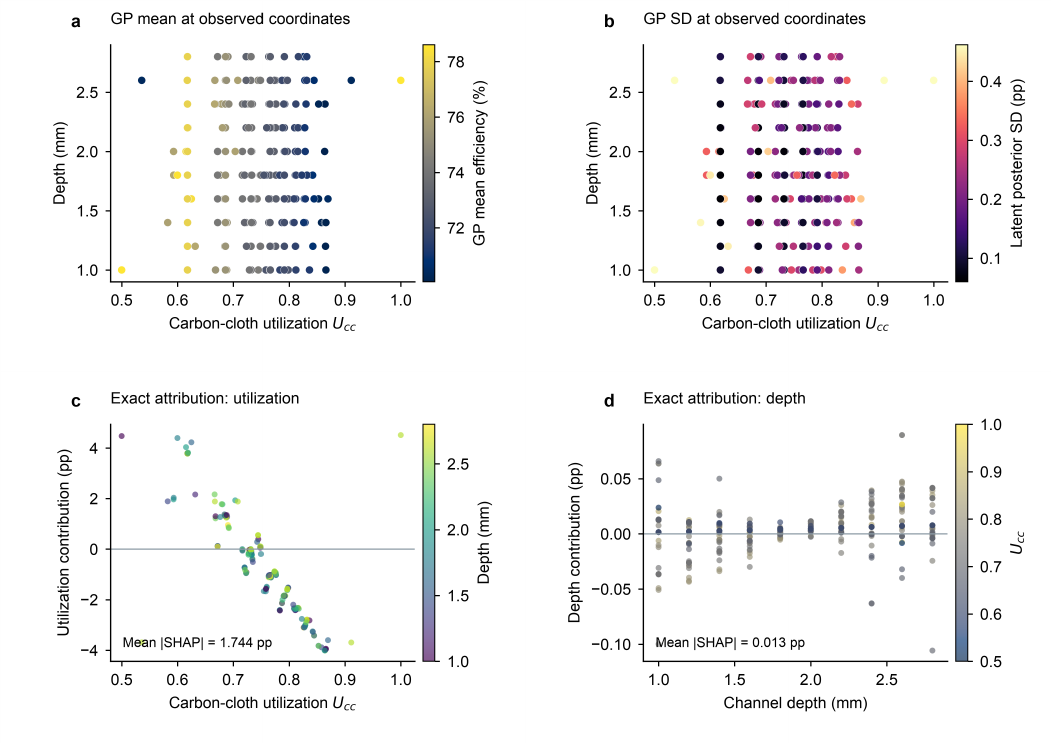}
\caption{Descriptive GP predictions and two-feature attribution. A separate GP is fitted to all 400 source records. a,b, Posterior mean and latent SD at the 329 observed descriptor pairs; unsampled regions remain blank. c,d, Exact interventional Shapley contributions using a fixed 100-record background, plotted at the original utilization and depth coordinates without jitter. All 400 contributions are retained, including overlap. The two vertical scales differ; mean absolute contributions provide a common-unit comparison. The attribution is conditional on interventional feature mixing, the background distribution and the fitted kernel.}
\label{fig4}
\end{figure}

\subsection{Robustness of candidate selection}

The 0.03970-point gap between the two highest responses motivates a practical target alongside exact recovery. All four GP policies reach within 0.05 percentage points of the original maximum in all 40 runs by evaluation 120, compared with 26 of 40 random runs (Figure 5a). In a sensitivity test that omits only record 400, GP-EI and GP-AL recover the remaining maximum of 78.7351\% in 20 of 20 paired runs, GP-UCB and GP-greedy in 19 of 20, and random search in 5 of 20 (Figure 5b). The GP advantage does not rest on record 400 alone.

We also examined cases with the same descriptor pair. The 25 duplicated coordinates contain 96 records with within-coordinate response ranges up to 1.856 percentage points (Figure 5c). The GP validation RMSE ranges from 0.705 to 0.772 across grouping labels rounded to 3, 4, 6, 8 and 9 decimal places, compared with 0.737 for exact groups (Figure S3). Changing the grouping resolution thus moves the RMSE by at most 0.035 percentage points. In the larger-initialization and coordinate-mean benchmarks, all four GP policies again recover the maximum in every run (Figure S3; Table S2).

\begin{figure}[htbp]
\centering
\includegraphics[width=\textwidth]{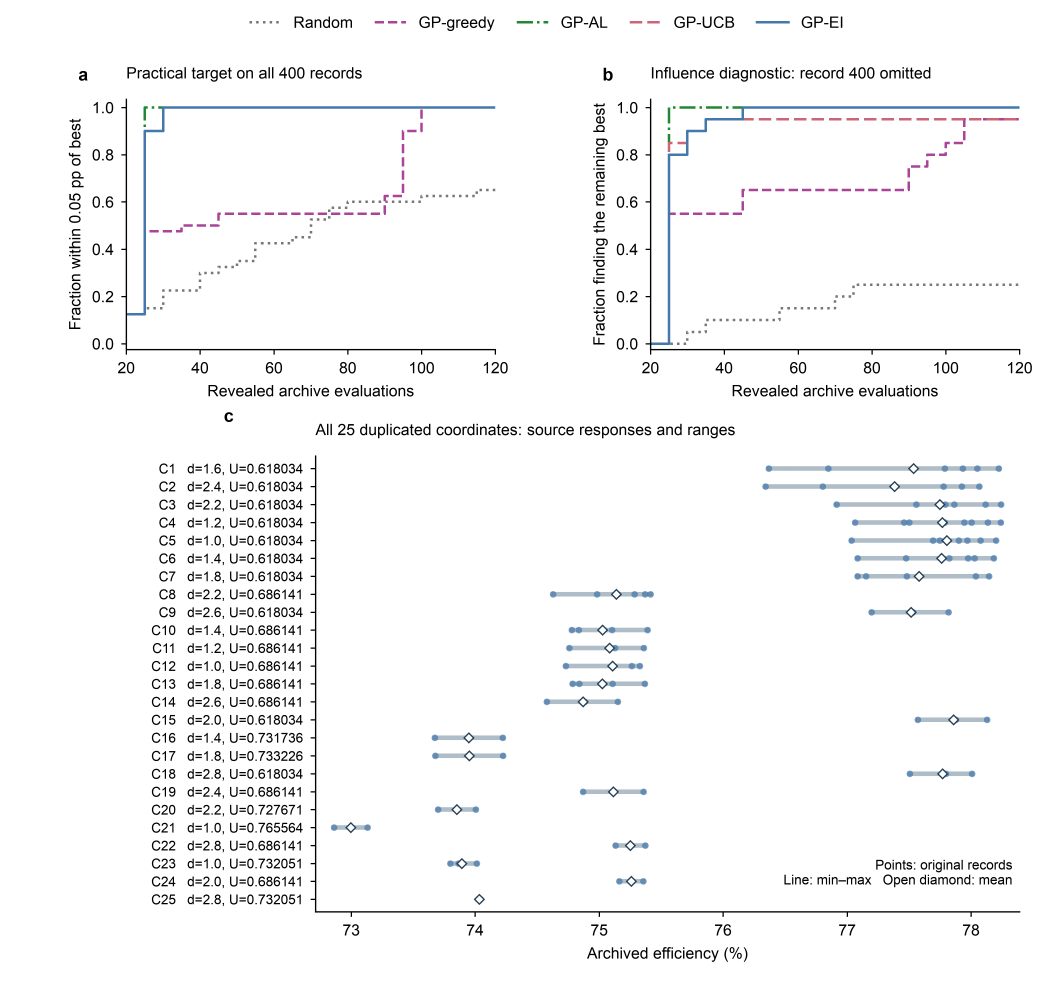}
\caption{Target sensitivity and descriptor multiplicity. a, Fraction reaching within 0.05 pp of the original maximum of 78.7748\%, across the 40 primary paired seeds. b, Post-hoc influence diagnostic omitting record 400: 399 candidates, 20 paired seeds, 20 initial records, batches of five and budget 120. The target becomes the remaining maximum of 78.7351\%. c, All 96 responses at the 25 duplicated descriptor pairs. Lines span the source minimum and maximum; open diamonds mark coordinate means. C1--C25 identify full-precision coordinates and source records in the audit table; printed utilization labels are rounded to six decimal places. Overlap is retained.}
\label{fig5}
\end{figure}

\subsection{Evaluating new flow-field geometries}

To evaluate new flow fields, FlowBD-A1 can use the COMSOL adapter in place of the archive evaluator. The adapter sets up and meshes each geometry and extracts the voltages, cell current and pumping power from the solution. Each candidate should retain its channel width, spacing and depth together with the utilization calculation. The isolated high-efficiency cases and the cases with shared descriptors are useful starting points for comparisons of flow and concentration fields. A broader design study could also consider cycle efficiency, degradation, lifetime and manufacturing cost.

\section{Conclusion}

In a replay of 400 archived simulations, FlowBD-A1 found the interdigitated flow field with the highest efficiency in the archive. GP-EI and GP-UCB recovered the archived maximum of 78.7748\% in all 40 paired runs within 120 evaluations, compared with 12 of 40 random-search runs. Median evaluation counts, including initialization, were 25 for uncertainty sampling, 30 for GP-EI and 95 for selection by the GP mean alone. Uncertainty sampling recovered the maximum in all 40 runs, compared with 39 of 40 for selection by the GP mean. When the boundary maximum was omitted, both GP-EI and uncertainty sampling recovered the remaining maximum in all 20 paired runs. Exploring sparsely sampled regions early paid off here because the highest efficiency lay outside the main group of high-efficiency designs.

Carbon-cloth utilization helped locate a range where high efficiencies were common. Cases with the same channel depth and utilization still gave different efficiencies, so the final comparison needs the full geometry. Comparing designs at the same operating point lets us examine the charge and discharge voltages separately from pumping power. For the best design, deep, wide channels and a narrow rib drive the same flow through the electrode as neighboring geometries at the lowest inlet pressure. FlowBD-A1 narrows the candidates to the few geometries worth this analysis, and the final choice weighs reactant supply to the electrode against the power needed to circulate the electrolyte.

\section*{Declaration of competing interest}

The authors declare that they have no known competing financial interests or personal relationships that could have appeared to influence the work reported in this paper.

\section*{Funding}

This work was supported by the Shanghai Basic Research Program on General Artificial Intelligence Large Models (2025SHZDZX025D02), the Young Talent Support Project of the Beijing High-Level Innovation and Entrepreneurship Talent Program (20250867), the Beijing Natural Science Foundation (2262028), the Young Scientists Fund of the National Natural Science Foundation of China (22308376) and the Interdisciplinary Frontier Exploration Program of China University of Petroleum, Beijing (2462025XKQY002).

\section*{Data and code availability}

The code, data and reproducibility archive are openly available in the gpbo directory of \url{https://github.com/cupzhoutianhang/FLOWBD-A1} (release tag gpbo-ces-2026-11-25) under the MIT License. This archive includes the original scalar tables, 500 archived-response optimization runs with 78,000 queries, grouped held-out predictions, coordinate-grouping diagnostics, two-feature Shapley contributions and numerical source data for the figures. The code comprises FlowBD-A1, the archive evaluator, model-validation scripts and a configurable MPh/COMSOL evaluator. Installation instructions, dependency versions, an executable archive example and automated numerical checks are provided with the code. The example can be run without COMSOL. The Supporting Information contains the complete numerical settings and additional diagnostics.

\section*{Appendix A. Supplementary data}
Supplementary data to this article: physical-model settings; archived geometry, mesh and physical-field panels; descriptor definition; computational protocol and complete benchmark results; sensitivity analyses; GP and comparison-model settings; adaptive evaluation algorithm; online output definitions. The Supporting Information is provided as a separate document.

\bibliographystyle{elsarticle-harv}

\end{document}